\documentstyle[aaspp4,tighten,flushrt]{article}

\begin{document}

\newcommand{\beq}{\begin{equation}}
\newcommand{\eeq}{\end{equation}}
\newcommand{\beqa}{\begin{eqnarray}}
\newcommand{\eeqa}{\end{eqnarray}}

\newcommand{\lexp}{\mathop{\langle}}
\newcommand{\rexp}{\mathop{\rangle}}

\def\eq#1{equation~(\ref{#1})}

\title{Large-scale Structure and the Redshift-Distance Relation}

\author{M. A. Jones and J. N. Fry}

\authoremail{mike@phys.ufl.edu, fry@phys.ufl.edu}

\affil{Department of Physics, University of Florida, %
Gainesville, FL  32611-8440; 
mike@phys.ufl.edu, fry@phys.ufl.edu}

\begin{abstract}

In efforts to demonstrate the linear Hubble law $ v = H r $ from 
galaxy observations, the underlying simplicity is often obscured 
by complexities arising from magnitude-limited data.
In this paper we point out a simple but previously unremarked fact: 
that the shapes and orientations of structures in redshift space 
contain in themselves independent information about the 
cosmological redshift-distance relation.

The orientations of voids in the CfA slice support the Hubble law, 
giving a redshift-distance power index $ p = 0.83 \pm 0.36 $ 
(void data from Slezak, de Lapparent, \& Bijoui 1993) or 
$ p = 0.99 \pm 0.38 $ (void data from Malik \& Subramanian 1997).

\end{abstract}

\subjectheadings{large-scale structure of universe}

\section{Introduction}

Hubble's (1929) observation that redshift increases linearly with 
distance for nearby galaxies has now been known for almost seventy 
years, and it is likely that its validity is not doubted by many.
Yet, any attempt to demonstrate this simple law from galaxy observations 
soon descends into the complexities of magnitude-limited observations,
the broad galaxy luminosity function, Malmquist bias, and, 
at larger redshifts, $K$-corrections, evolution, etc.
This has permitted an often-ignored but persistent challenge to 
a linear redshift law from those preferring a quadratic relation, with 
both challengers and supporters citing data ranging from relatively 
nearby, bright optical galaxies (Soneira 1979, Segal 1980) through the 
{\it IRAS} 1.2~Jy redshift catalog 
(Segal et al. 1993; Koranyi \& Strauss 1997).

One complication in understanding the classical Hubble diagram has 
been inhomogeneity, appearing as a dependence of density on radius 
or redshift or in clustering and peculiar velocities.
With the advent of more galaxy redshift catalogs covering larger
amounts of the sky, we obtain ever clearer pictures of the universe.
The CfA ``slice'' (de Lapparent, Geller, \& Huchra 1986), in particular,
first revealed dramatic structures, large voids separated by
well-defined walls, extending to a significant fraction of the survey
volume, perhaps even calling into question whether a survey to this
depth represents a statistically fair sample of the universe.
In other ways, however, this inhomogeneity itself can be useful.
In this paper we use the shapes and orientations of structures 
to obtain information about the background 
cosmology in which they are embedded. 
In Section 2 below we discuss how shapes and orientations of 
structures in redshift space depend on the redshift-distance relation, 
and in Section 3 we apply these considerations to 
the CfA slice (de Lapparent, Geller, \& Huchra 1986) and the
Las Campanas redshift survey (Shectman et al 1996, LCRS).
Section 4 contains a final discussion.

\section{Structure, Homogeneity, and the Redshift-Distance Relation}

The usual Hubble law $ cz = Hr $ relating cosmological redshift
$z$ and distance $r$ is the linear case of a power-law relation
characterized by an index $p$,
\beq
z = A r^p . \label{Arp}
\eeq
(As in the usual relation derived from cosmological Robertson-Walker 
solutions to the Einstein equations, this behavior can be the small-$r$ 
limit of a possibly more complex relation.)
Since flux decreases with distance as $ r^{-2} $, the classical
Hubble diagram uses recession velocities and apparent magnitudes to
infer $p$ from the magnitude-redshift relation, 
\beq
m = M + {5 \over p} \log {z \over z_0} .
\eeq
However, this simple relation is difficult to observe in practice
because of the broad distribution of galaxy luminosities.

We present an alternative method making use of inhomogeneities 
to determine $p$ from the shapes and orientations of structures 
in redshift space.
Shapes and orientations in redshift space have been considered
previously as indications of cosmological effects that depend
on the density parameter $ \Omega_0 $, the deceleration parameter
$ q_0 $, or the cosmological constant $ \Lambda $,
using as tracers the distribution of galaxies in clusters
(Alcock \& Paczy\'{n}ski 1979), the angle with the line of sight of
the line joining neighboring pairs of quasars (Phillipps 1994),
or distortions of voids (Ryden 1995).
Because these effects are weak at small redshift,
with typical behaviors of the form $ 1 + k z $ for some $k$, 
implications for cosmological parameters have been inconclusive.
However, the effect of a redshift index $ p \ne 1 $ appears at zero order.

For example, suppose that we can identify features with a
characteristic angular extent $ \Delta \theta $ and redshift extent
$ \Delta z $.
From the redshift-magnitude relation of \eq{Arp}, 
$ \Delta \theta $ is related to the transverse physical diameter of 
the structure $ r_\perp $ by $ \Delta \theta = \Delta r_\perp / r $; 
while $ \Delta z $ is related to the line-of-sight diameter
$ \Delta r_\parallel $ by $ \Delta z/z = p \, \Delta r_\parallel / r $
($ \Delta r \ll r $).
Thus, without ever needing to know the distance $r$, we have
\beq
{1 \over z} {{\Delta z} \over {\Delta \theta}} =
p \, {{\Delta r_\parallel} \over {\Delta r_\perp}} ;
\eeq
the aspect ratio in redshift space is related to the aspect ratio
of the physical object by exactly the redshift index $p$.
Ignoring peculiar velocities, a (distant) sphere is mapped in 
redshift space into an ellipse stretched by a factor $p$
along the line of sight.

With irregularly shaped objects, sparsely sampled, 
this aspect ratio may be difficult to measure accurately.
Another quantitative measure of $p$ can be obtained from 
considering orientations.
The line joining the center of a sphere to a point on the surface
makes an angle $ \theta $ with the line of sight 
(this is always an angle between 0 and $ 90 \arcdeg $).
Consider the average of the cosine of this angle in redshift space, 
$ \mu = \lexp \cos \theta \rexp $.
In three dimensions this is simply
\beq
\mu_{3d} = {p \over p+1} . \label{mu3d}
\eeq
Inverting, the redshift index $p$ is found from $ \mu $
as $ p = \mu / (1-\mu) $, with error
$ \Delta p = \sigma_\mu / (1-\mu)^2 $.
For a two-dimensional sample such as the CfA slice,
the expression is a bit more complicated, 
\beq
\mu_{2d} = {2 \over \pi} \sqrt{ p^2 \over p^2-1}
\tan^{-1} \sqrt{p^2 - 1} \label{mu2d}
\eeq
for $ p > 1 $, analytically continued for $ p < 1 $,
with $ \mu_{2d} = 2 / \pi $ and $ d \mu / dp = 2/3\pi $ at $ p = 1 $.
For all $p$ we have $ \mu_{2d} > \mu_{3d} $, but the basic behavior
remains the same, $\mu$ is a smooth, monotonic function of $p$.
Figure~1 shows $\mu$ as a function of $p$ for both two and
three dimensions.
These expressions hold not only for physical objects that are 
inherently round, but also for ellipses of any eccentricity, 
provided only that orientations to the line of sight are random.
Equation~(\ref{mu2d}) will allow us to determine $p$ from data.

\section{Application to Observations}

We first present the effect visually using the Las Campanas 
Redshift Survey (Shectman et al. 1996), comprising in total
26,418 redshifts in six $ 1\fdg 5 \times 80 \arcdeg $ strips in 
selected north and south pole areas with median redshift of 
approximately $ 30,000 \, {\rm km} \, {\rm s}^{-1} $.
Figure~2a shows the LCRS $-12^\circ$ declination slice in 
angle-redshift space, which is also physical $\theta$-$r$ space 
if distance and redshift are related by the linear Hubble law.
Represented in the figure are $4071$ LCRS galaxies with $cz$
less than $ 45,000 \, {\rm km \, s}^{-1} $.
Figure~2b shows how the slice would look in $\theta$-$r$ 
if $ r \sim z^2 $ ($ p = \case12 $ in eq.~[\ref{Arp}]; 
this is also how the slice would look in redshift space if
true distances are as in Figure~2a and $ p = 2 $).
Figure~2c shows physical distances if $ r \sim z^{1/2} $
($ p = 2 $; this is how the slice would look in redshift space if 
distances are as in Figure~2a and $ p = \case12 $).
In Figure~2a we see some structures that are elongated radially,
and some that are elongated in the transverse direction, as we 
expect if the universe is homogeneous and we do not live 
in a privileged location.
In Figure~2b, however, essentially all elongations are radial,
and in Figure~2c, essentially all are transverse.
Further, only for $ p = 1 $ is the density not steeply 
varying with $r$.

We can attempt to put this on a quantitative basis by examining
nearest neighbor orientations in the CfA ``slice of the universe'' 
(de Lapparent, Geller, \& Huchra 1986), 1093 galaxies brighter 
than 15.5 in magnitude in the $ 6 \arcdeg $ declination band 
$ 25\fdg 5 < \delta < 32\fdg 5 $ covering $ 135 \arcdeg $ in 
right ascension from $ 8^{\rm h} $ to $ 17^{\rm h} $.
For the 1057 galaxies in the inner 80\% of the CfA slice,
the average of the orientation of the line 
joining nearest neighbors gives $ \mu = 0.644 \pm 0.010 $.
From eq.~(\ref{mu2d}), this corresponds to $ p = 1.037 \pm 0.050 $.
This appears to support the Hubble law, but in fact is only 
a reflection of the metric we have used to determine 
``nearest neighbor'' in redshift space,
$ \Delta s^2 = \Delta z^2 + z^2 \Delta \theta^2 $.
If we first square all the redshifts and repeat the analysis 
we obtain $ \mu = 0.670 \pm 0.010 $; if we take the square root 
of all redshifts we obtain $ \mu = 0.634 \pm 0.010 $.
The explanation is simple: the radial rescaling changes 
which points are nearest neighbors in redshift space.
This may at first be surprising, but seems obvious after considering 
the following.
Take a uniform distribution of points in the $x$-$y$ plane and then
stretch it by an arbitrary amount along the $x$-axis.
The resulting distribution was, and thus still is, independent of $y$; 
and it was, and so still is, independent of $x$; 
and although the overall density may be changed, the distribution 
must again be uniform in the portion of the plane it occupies.
Thus we conclude that nearest neighbors do not provide the quantitative
measure of $p$ we seek.

We need to use objects whose identity is not changed by remapping the 
radial coordinate, and an obvious candidate is voids.
The identification of a void relies only on distinguishing locally 
underdense regions and not on an absolutely calibrated sensitivity.
Isolated voids become innately round; unlike collapsing objects, which
typically collapse along one axis first, non-spherical voids become
more spherical as they evolve (Bertschinger 1985; Ryden 1994).
Even if a typical bubble is not perfectly round in shape, from
homogeneity we expect the distribution of shapes to be independent
of $z$, and from isotropy we expect orientations to be random.
We use the orientations of voids identified in two wavelet analyses 
of the CfA slice to determine the redshift-distance index $p$.
Slezak, de Lapparent, \& Bijoui (1993) present a list of 15 voids, 
identified using an isotropic template with orientation defined 
by fitting to an ellipse after their wavelet analysis detects 
a significantly underdense region.
Malik \& Subramanian (1997) list a total of 29 voids, from which we 
remove three apparent duplications and one that is perfectly round, 
so that no orientation is determined.
Their anisotropic wavelet templates includes orientation as a parameter, 
but their angular resolution is limited to multiples of $ 30 \arcdeg $.
Results are given in Tables 1 and 2.
The best determinations are those averaged over all 15 Slezak voids, 
$ \mu = 0.597 \pm 0.0941 $ corresponding to $ p = 0.831 \pm 0.357 $; 
and the $ 7\sigma $ significance level from Malik \& Subramanian, 
which gives $ \mu = 0.635 \pm  0.0807 $, or $ p = 0.992 \pm 0.377 $.
Values for various subsets are given in Table~1 and Table~2.
If we look for the best integer index, these results prefer 
$ p = 1 $ to $ p = 2 $.

We might expect peculiar velocities to distort the apparent shapes
of expanding voids.
Voids generally expand more rapidly than the Hubble flow, 
but the expansion velocity of the shell of material around a hole 
typically is only a small fraction of the Hubble velocity
across the hole (Peebles 1982; Ryden 1994).
For isolated, spherical ``top-hat'' voids we can compute the expansion 
velocity, typically expressed as a fraction $F$ of the Hubble flow.
For $ \Omega = 1 $, this factor is $ F = \case13 $ if the void is
uncompensated and $ F = \case15 $ if compensated (Bertschinger 1985), 
which becomes $ F = \case15 \Omega^{0.6} $ for $ \Omega \la 1 $ 
(Reg\H{o}s \& Geller 1991); however, the flow at a void boundary 
generally depends critically on its surroundings.
Measured differences between redshift and independent Tully-Fisher
distance determinations for galaxies in the largest bubble of the CfA
slice confirm that peculiar velocities are small, consistent with zero
and with a 1-$ \sigma $ upper bound less than 5\% of the diameter 
(Bothun et al. 1992; this determination of course assumes $ p=1 $).
It is instructive to compare with the results for overdense regions, 
also provided by Malik \& Subramanian (1997).
As a whole, overdense detections are smaller and more anisotropic than 
are the underdense regions, and for the higher significance level 
detections, they are preferentially oriented along the line of sight, 
with $ \mu = 0.827 \pm 0.066 $ for significance $ > 10 \sigma $.
This is as we expect; collapsed objects are smaller and have larger 
internal velocity dispersion.

\section{Discussion}

We have shown in this paper how the apparent shapes and 
orientations of objects in redshift space can be used to 
determine the redshift-distance relation.
When redshift is not linear in distance, objects such as voids that 
are intrinsically round in space when viewed in redshift space are 
stretched along the line of sight; and an initially isotropic 
distribution of orientations becomes distorted in the radial direction, 
with measurable effect, as in \eq{mu2d}.
Application to voids in the CfA slice gives redshift power index 
$ p = 0.83 \pm 0.36 $ from Slezak et al. (1993) and 
$ p = 0.99 \pm 0.38 $ from Malik \& Subramanian (1997), 
both a modest preference for $ p = 1 $ over $ p = 2 $.
To obtain these results in redshift space with a true $ p = 2 $, 
voids in the CfA slice in space would have to be flattened and 
preferentially aligned transverse to the line of sight.
The main complication to the simple interpretation is likely 
to come from peculiar velocities; an expansion velocity will 
introduce an apparent line-of-sight elongation.
However, this effect is expected to be small: 
in the most extreme case, of an uncompensated, completely empty, 
isolated void in an $ \Omega_0 = 1 $ universe, expansion would 
make $ p = 1 $ appear to be $ p = \case43 $.
In any case, unless voids are contracting, a condition difficult 
to make physical sense of, peculiar velocities will only increase, 
not decrease, the apparent value of $p$, and will not distort  
a quadratic redshift law to appear to prefer $ p = 1 $.

We do not expect void orientation statistics to replace 
classical methods for demonstrating the Hubble law.
The most precise confirmation of the linearity of the Hubble law, 
$ p/5 = 0.2010 \pm 0.0035 $, has been obtained recently from the 
Hubble diagram of type Ia supernovae (Riess, Press, \& Kirshner 1996).
The luminosity function as a function of redshift and 
the radial dependence of galaxy density in the {\it IRAS} 1.2-Jy catalog 
(Koranyi \& Strauss 1997) also support the linear Hubble law.
However, our method offers an independent verification of the Hubble 
law in which magnitude-limited data, Malmquist bias, and evolution 
do not present problems.
Further examination of the effect in real data, especially including 
in detail the peculiar velocities of the void walls, 
will undoubtedly introduce complications to the simple relations 
in equations (\ref{mu3d}) and (\ref{mu2d}).
Whatever these complications, they will be different from those
that enter arguments over the redshift-magnitude relation.

\acknowledgments

We thank MJG for providing the CfA slice data in machine readable form.
This research was supported in part by NASA grant NAG5-2835.

\clearpage

\begin{deluxetable}{crcc}
\tablecaption{Redshift Index $p$ from Slezak et al. (1993) Voids}
\label{table1}
\tablehead{
\colhead{Void size $ ({\rm km \, s}^{-1}) $} & \colhead{$N$} &
\colhead{$ \mu $} & \colhead{$ p $} }
\startdata
  350--700  & 7 & $ 0.594 \pm 0.130 $ & $ 0.821 \pm 0.488 $ \nl
 1000--2000 & 8 & $ 0.598 \pm 0.143 $ & $ 0.836 \pm 0.547 $ \nl
 350--2000 & 15 & $ 0.597 \pm 0.094 $ & $ 0.831 \pm 0.357 $ \nl
\enddata
\end{deluxetable}

\begin{deluxetable}{crcc}
\tablecaption{Redshift Index $p$ from Malik \& Subramanian (1997) Voids}
\label{table2}
\tablehead{
\colhead{Significance} & \colhead{$N$} & 
\colhead{$ \mu $} & \colhead{$ p $} }
\startdata
$ > 6 \sigma $ & 22 & $ 0.685 \pm 0.070 $ & $ 1.265 \pm 0.477 $ \nl
$ > 7 \sigma $ & 18 & $ 0.635 \pm 0.081 $ & $ 0.992 \pm 0.377 $ \nl
$ > 8 \sigma $ &  8 & $ 0.654 \pm 0.122 $ & $ 1.086 \pm 0.634 $ \nl
\enddata
\end{deluxetable}

\clearpage

\plotone{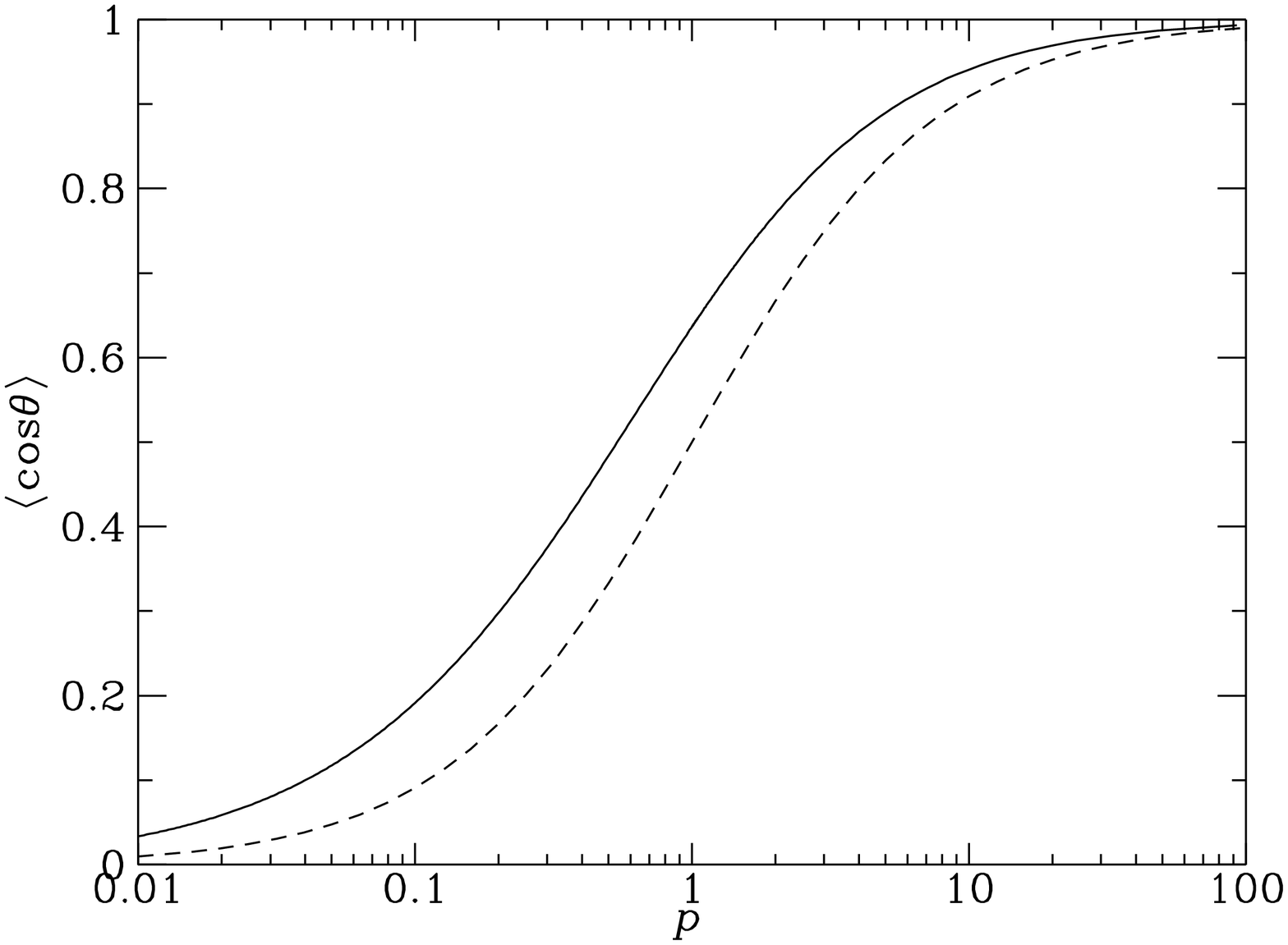}
\figcaption
{The average $ \mu = \lexp \cos \theta \rexp $ over orientation 
relative to line of sight in redshift space for an isotropic 
distribution of orientations in space, 
as a function of the redshift index $p$.
The dashed line shows the result in three dimensions, the solid line
the result for a two-dimensional slice. \label{fig1}}

\plotone{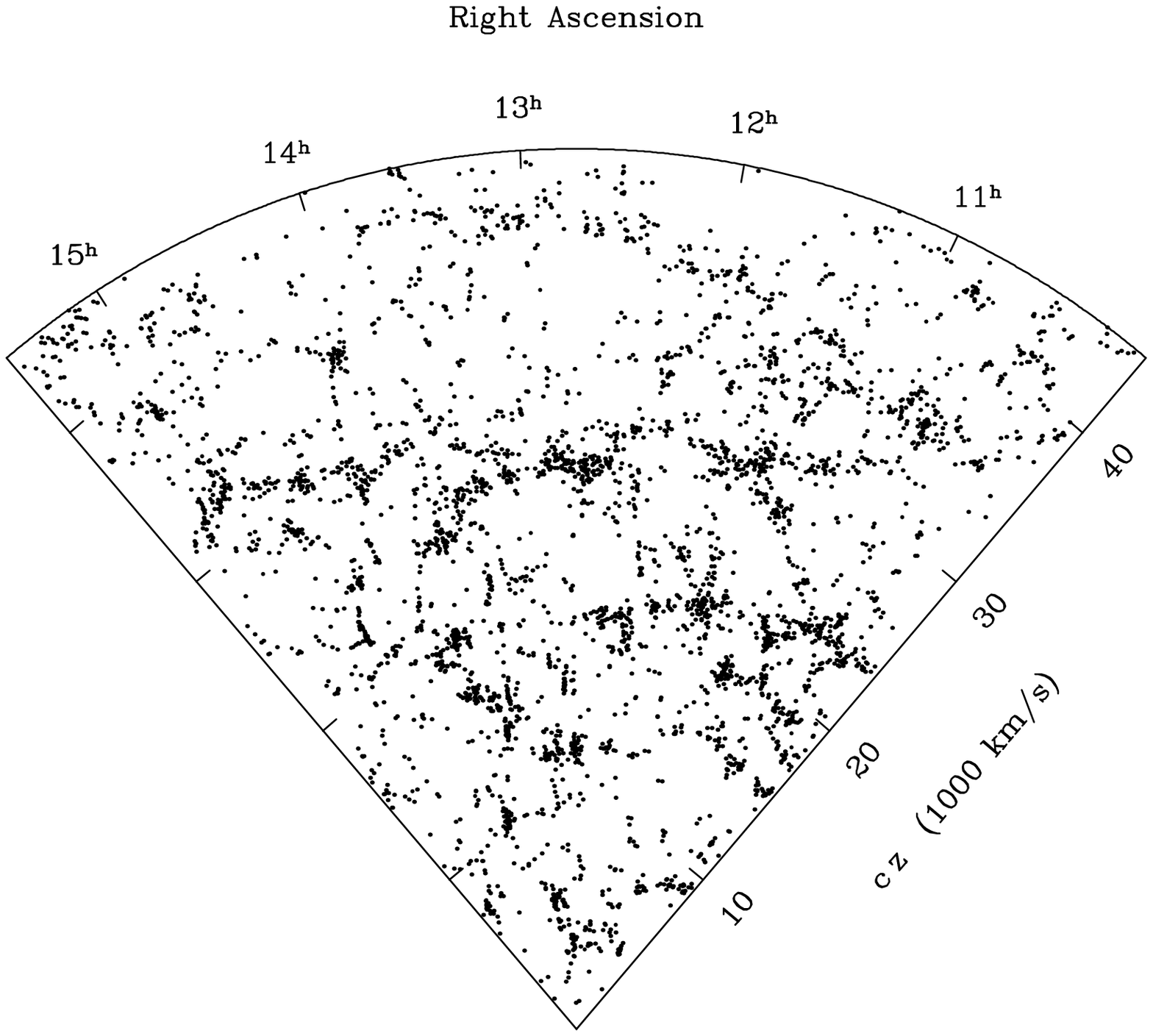}
\figurenum{2a}
\figcaption
{The LCRS galaxy distribution in $z$ and $\theta$.\label{2a}}

\plotone{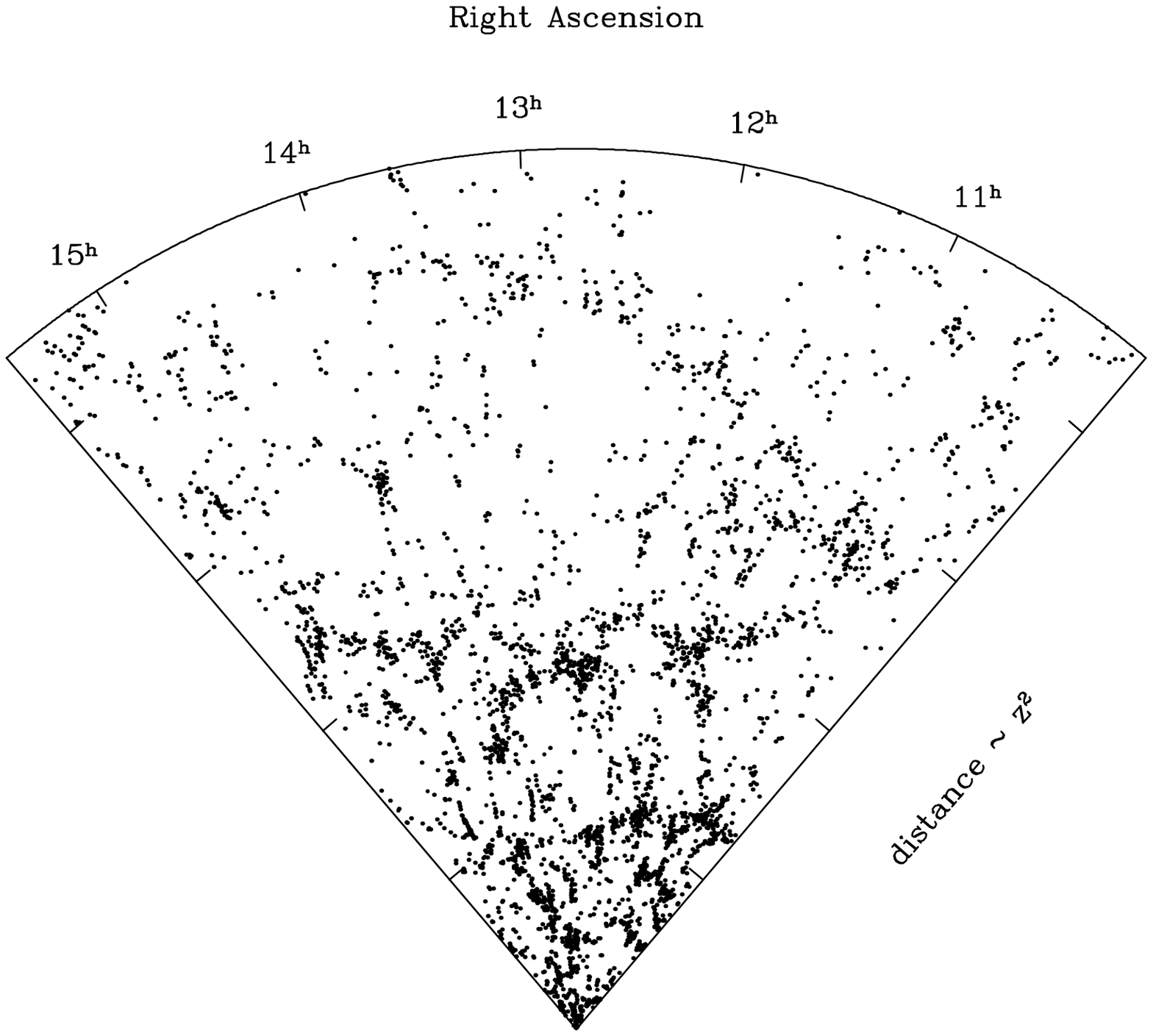}
\figurenum{2b}
\figcaption
{The LCRS galaxy distribution in $r$ and $\theta$ 
for $ z \sim r^{1/2} $, or $ r \sim z^2 $.
\label{2b}}

\plotone{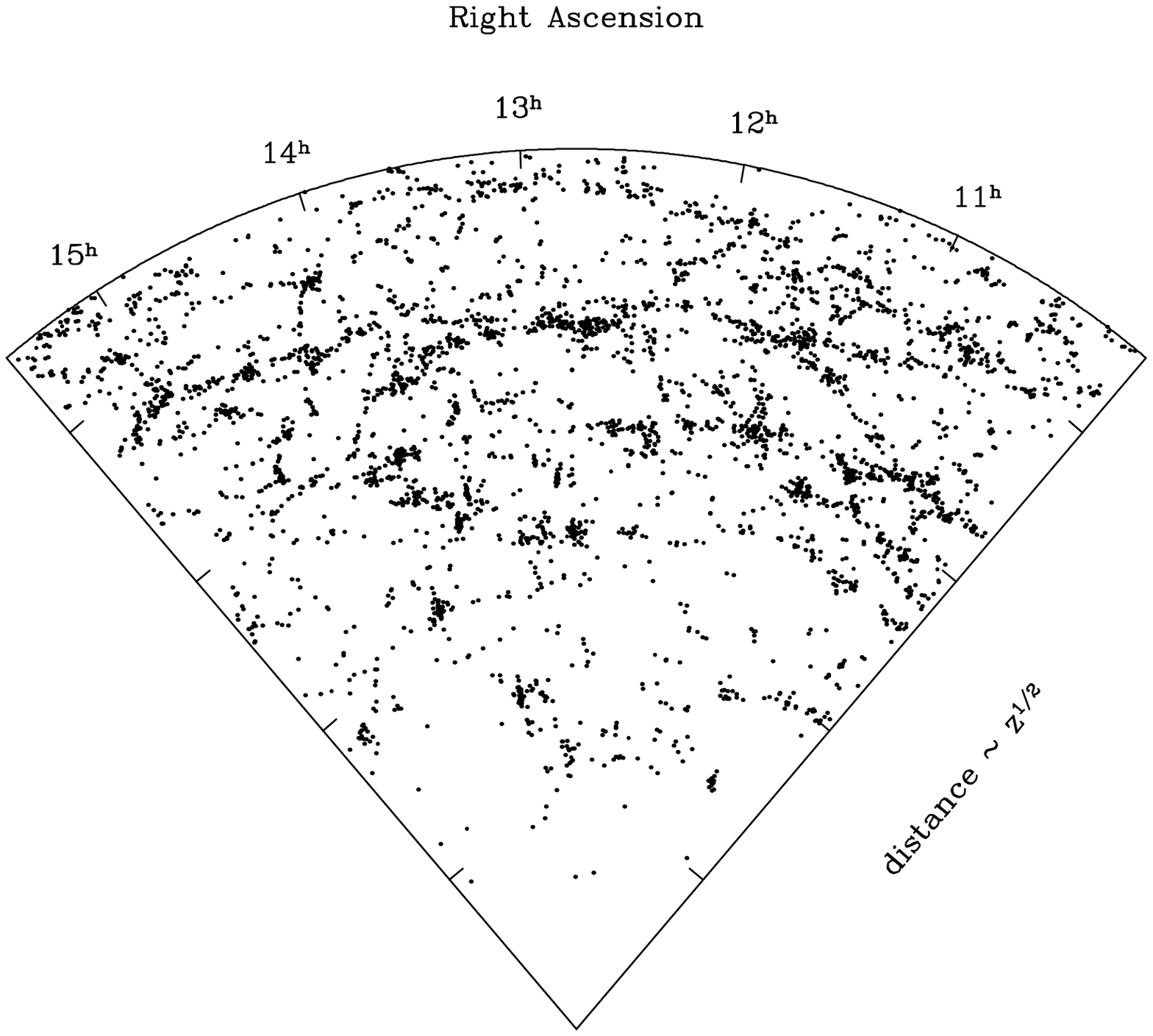}
\figurenum{2c}
\figcaption
{The LCRS galaxy distribution in $r$ and $\theta$ 
for $ z \sim r^{2} $, or $ r \sim z^{1/2} $.
\label{2c}}


\clearpage


\begin{references}

\reference{AP79}
Alcock, C., \& Paczy\'{n}ski, B. 1979, Nature, 281, 358

\reference{B85}
Bertschinger, E. 1985, \apjs, 58, 1

\reference{BGKHS92}
Bothun, G. D., Geller, M. J., Kurtz, M. J., Huchra, J. P., \&
Schild, R. E. 1992, \apj, 395, 347    

\reference{CfA}
de Lapparent, V., Geller, M.~J., \& Huchra, J.~P. 1986, ApJ, 302, L1

\reference{H29}
Hubble, E.  1929, Proc NAS, 15, 168

\reference{KS97}
Koranyi, D. M., \& Strauss, M. A. 1997, \apj, 477, 36

\reference{MS97}
Malik, R. K., \& Subramanian, K. 1997, \aap, 317, 318

\reference{P82}
Peebles, P.~J.~E. 1982, \apj, 257, 438

\reference{P94}
Phillipps, S. 1994, \mnras, 269, 1077

\reference{RG91}
Reg\H{o}s, E. \& Geller, M. J. 1991, \apj, 377, 14

\reference{RPK96}
Riess, A. G., Press, W. H., \& Kirshner, R. P. 1996, \apj, 473, 88

\reference{R95}
Ryden, B. 1994, \apj, 423, 534

\reference{R95}
Ryden, B. 1995, \apj, 452, 25

\reference{S80}
Segal, I. E. 1980, \mnras, 192, 755

\reference{SNWZ93}
Segal, I. E., Nicoll, J. F., Wu, P., \& Zhou, Z. 1993, \apj, 411, 465

\reference{LCRS}
Shectman, S. A., Landy, S. D., Oemler, A., Tucker, D. L., Lin, H.,
Kirshner, R. P., \& Schechter, P. L. 1996, \apj, 470, 172

\reference{DsLB93}
Slezak, E., de Lapparent, V., \& Bijoui A. 1993, \apj, 409, 517

\reference{S79}
Soneira, R. P. 1979, \apj, 230, L63

\end{references}
\end{document}